\documentclass{elsart}%
\usepackage{amsfonts}
\usepackage{natbib}

\usepackage{amsmath}
\usepackage{amssymb}

\begin{document}
\begin{frontmatter}
\title{Equilibrium of stellar
dynamical systems in the context of the Vlasov-Poisson model}
\author{J\'er\^{o}me Perez}
\address{Applied Mathematics Laboratory, Ecole Nationale Sup\'erieure de Techniques Avanc\'ees, 32 Bd Victor, 75739 Paris Cedex 15}
\ead{jerome.perez@ensta.fr}
\begin{abstract}
This short review is devoted to the problem of the equilibrium of stellar
dynamical systems in the context of the Vlasov-Poisson model. In a first part
we will review some classical problems posed by the application of the
Vlasov-Poisson model to the astrophysical systems like globular clusters or
galaxies. In a second part we will recall some recent numerical results which
may give us some some quantitative hints about the equilibrium state  associated to those
systems.\end{abstract}
\end{frontmatter}

\section{Introduction}
Globular clusters and galaxies are concentration of stars whose physical characteristics are such that allow astrophysicists to model them as being in some equilibrium state for not too long time scales. The large number of stars and some properties of the gravitational interaction $-$ which evidently play a key role for their dynamics $-$ made the Vlasov-Poisson model a good candidate for their modeling. In this short review we present some results and problems of this field of research.

\section{The Vlasov-Poisson system in the context of stellar dynamics\label{sect2}}
The Vlasov-Poisson model is a mean field approximation of the gravitational potential generated by a large assembly of point masses in the context of the non dissipative kinetic theory. 
In order to apply this model to self-gravitating systems like
globular clusters or galaxies, we have to pose three classical hypotheses :
First of all, it suffices to consider only Newtonian gravity for the
mean field interaction between stars
\footnote{The general theory of relativity could also be considered in the context of the
cosmological principle. The system takes then into account the scale factor $%
a\left( t\right) $ of the Universe. This formalism is often used in the context of formation of the large structures of the Universe.}. Second, all stars are statistically equivalent, particularly they do have the same
mass and they do not evolve. This is a drastic approximation but it could be
relevant in a mean sense and during quite long evolution stages of these objects.
Third, the considered systems must be non-dissipative on the time scales of interest. 
Nevertheless, it is well known that stellar dynamical systems could be dissipative
mainly by two processes : On the one hand , such systems could contain gas
which dissipates their total energy by dynamical friction. This feature must
be considered for modeling the visible part of spiral galaxies.
On the other hand it is well known that deviations from the mean-field forces due to the discreteness of the
    actual stellar distribution will alter the distribution of kinetic
    versus potential energies as computed from the Vlasov-Poisson
    equations. 
    In fact, this is a problem of time scales. The dynamical
time $T_{d}$ of self-gravitating systems depends only on its mean density $%
\tilde{\rho}$ via the relation$\ T_{d}\approx 1/\sqrt{G\tilde{\rho}}$ where $G$ is the Newton gravitational constant (see \cite
{BT87} for details). It is typically the time taken by a test star to cross
the system. Another duration is under interest : the time $T_{r}\ $required
for the mean velocity of the system to change by of order itself. This time has been estimated by Chandrasekhar for uniform systems to
    be proportional  to the dynamical one via the relation $T_{r}\ \approx N\
T_{d}/\ln N$ (see \cite{BT87} p. 189 for details). It is typically the time for which the dissipative process like encounters plays a key role in the system's dynamics.  Galaxies $\ $%
are composed of at least $N=10^{9}$ stars, therefore this time limitation is not
relevant in this case. The situation is not so clear for globular clusters.  The number of their components ranges from  $10^{4}$ to $10^{6}$. If initial stages (during a few
hundred of dynamical times ...) could be non dissipative and therefore
potentially described by Vlasov equation, during late stages $-$ for billions years
clusters $-$ encounters dissipate energy and stellar dynamicists introduce
Fokker-Planck formalism.

\section{Equilibrium as steady state solutions of the Vlasov-Poisson system}
Self-gravitating collisionless systems could be modelized by a phase space distribution function $f\left(\mathbf{r},\mathbf{p},t\right)$ and a mean field potential $\psi \left(\mathbf{r},t\right)$. Vectors $\mathbf{r}$ and $\mathbf{p}$ are the usual conjugated position and momentum which are elements of $\mathbb{R}^{d}$ for $d$ dimensional systems. The functions $f$ and $\psi$ are coupled by the Vlasov-Poisson system, which is for $d=3$:
\[
\left\{ 
\begin{array}{ccc}
\displaystyle \frac{\partial f}{\partial t}+\left\{ \;E\;,\;f\;\right\} =0 & \mbox{where }
& E=\displaystyle \frac{p^{2}}{2m}+m\psi \\ 
&  &  \\ 
\psi \left( \mathbf{r},t\right) =-Gm\displaystyle \int \frac{f\left( \mathbf{r}^{\prime
},\mathbf{p}^{\prime },t\right) }{\left| \mathbf{r}-\mathbf{r}^{\prime
}\right| }d\mathbf{r}^{\prime }d\mathbf{p}^{\prime } & \Leftrightarrow & 
\left[
\begin{array}{lc}
& \rho \left( \mathbf{r},t\right)\\
\Delta \psi =4\pi G & \overbrace{m\int f\left( \mathbf{r},\mathbf{p}%
^{\prime },t\right) d\mathbf{p}^{\prime }}  \\
& \hspace{-3cm} \ \psi \mbox{ bounded at }\infty
\end{array}\right.
\end{array}
\right. 
\]
The quantity $E$ denotes the mean field energy of a test star and plays a
central role in this system. The function $\rho \left( \mathbf{r}%
,t\right) \;$  represent the mean mass density distribution of the system. 
Writing the Vlasov equation as above $-$ i.e. using the Poisson brackets $-$ makes
trivial the well known result that every positive and normed function\ $f_{o}
$ of the mean field energy is a steady state solution of this system. A
natural question, initially posed by S.\ Chandrasekhar in the middle of the
last century, could then be : What are the properties of physical systems
whose distribution function writes $f_{o}=f_{o}\left( E\right)$ ? Although
the answer of the inverse problem  was well known by astrophysicists
\footnote{As a matter of fact, a spherical self-gravitating system must be associated to a radial gravitationnal potential which produces naturally a radial density profile by Poisson equation}, the natural direct one
was  solved only fifty years later \cite{PA96}, mainly by using in
this context a difficult, but classical, mathematical theorem by Gidas,
Ni and Nirenberg \cite{GNN81}. The main ingredients of
this result could be sketched as follow : If the distribution function
depends only on the mean field energy, then the mean mass density depends on
the position $\mathbf{r}$ only through the mean field potential
\[
f_{o}=f\left( E\right) \Rightarrow \rho _{o}\left( \mathbf{r}\right) =m\int
f\left( \frac{p^{\prime 2}}{2m}+m\psi _{o}\right) d\mathbf{p}^{\prime
}=\rho _{o}\left( \psi _{o}\right) 
\]
In this case, Poisson equation writes $\Delta \psi _{o}=c\;\rho _{o}\left(
\psi _{o}\right) \,\;$where $c$ is a positive constant. The physical context allows additional hypotheses : Newtonian gravity imposes
that $\psi _{o}\left( \mathbf{r}\right) $ is a negative function bounded at
infinity. The classical continuous limit let us consider that $\rho _{o}\left( \mathbf{r}%
\right) $ is a positive and continuous function. The GNN theorem \cite{GNN81} then allows us to claim that $\psi _{o}=\psi
_{o}\left( \left| \mathbf{r}\right| \right) $ and therefore
the  corresponding system has spherical symmetry in the spatial part
of the phase space. Properties of the system in the velocity part of the 
phase space are more evident : The dispersion velocity tensor is clearly
proportional to unity, therefore the system is said isotropic in velocity space.
Spherical and isotropic steady state solutions of Vlasov-Poisson systems was
intensively studied in the context of stellar dynamics as classical models
for globular clusters and galaxies. Stability of such systems was proved for
monotonic distributions functions in the linear case (see \cite{PA96} for
a whole presentation of this topic) and also in some general non linear
cases \cite{R05}.

Considering other  potential isolating integrals of motion  in the
gravitational field, we can define other classes of steady state solutions of
Vlasov-Poisson systems : This classical property of Poisson brackets is
sometimes called Jeans Theorem in stellar dynamics (see \cite{BT87}
for instance).

When the distribution function depend additionally on the mean angular momentum
modulus $f_{o}=f_{o}\left( E,L^{2}\right) $, an extension of the GNN theorem shows that if the mass density is monotonic, then the system is
always spherical. However, the tangential velocity dependence of $L^{2}=r^{2}v_{t}^{2}$
 makes the system anisotropic in the velocity space. The asymmetry of such
systems is associated to an imbalance in the ratio of radial over tangential star orbits velocity distribution. Since the end of the 80's, stellar dynamicists have understood that such
anisotropy could be at the origin of an interesting instability. As a matter
of fact, if anisotropic spherical stellar systems are generally stable against radial
perturbations , it could be proved that systems whose cannot be infinitesimally perturbed by
radial disturbances are intrinsically unstable\cite{PAAS96}. This is the fine
mechanism of Radial Orbit Instability : For pure radial orbit system, each star orbit extension is
exactly the radius of the whole system.  Thus, such a system \emph{cannot} receive infinitesimal radial
perturbation which affects by definition only an infinitesimal part of the
system. On the contrary, any non radial perturbation associated to a given
spatial direction could stretch or compress infinitesimally the spatial
extension of an associated star orbit. This feature generates a tidal
friction which makes an instability to grow and forms a triaxial system from an
initial sphere. Radial Orbit Instability triggers when a sufficient
amount of radial orbits is present in the system $-$ a general criterion is
given in \cite{PAAS96} from distribution function susceptibility to receive
radial perturbations. As indicated by numerical analysis \cite{RP04}, this
feature could be at the origin of triaxiality in some self-gravitating
systems like elliptical galaxies.

Less is known about steady state solutions characterized by distribution
functions depending, in addition of $E,$ on more complicated integrals.
If some special models associated to $f_{o}=f\left( E,L_{z}\right) $ are
clearly triaxial (see \cite{BT87} for details) , there is, up to now,  no
extensions of the famous GNN theorem which allows to claim anything in
a general way. 

\section{Equilibrium of a stellar system}

Taking into account the time problem limitation presented in section \ref
{sect2} and their generally quiet physical properties, one can modelize
 globular clusters and at least elliptical galaxies by steady state solutions of
the Vlasov-Poisson system. The fundamental question that we have to answer is : What are the associated distribution functions ? 

This problem was attacked by stellar dynamicists using three approaches : Thermodynamics, comparison with observational data  and numerical simulation.

The thermodynamical approach, e.g. \cite{emden}, is based on the classical assumption of
statistical physics which associates the equilibrium state to the maximum Boltzmann
entropy one. In the late sixties, \cite{LB67} reconsider this problem  and failed thoroughly : The classical result of these
works is the isothermal sphere which distribution function is $f_{o}=f\left(
E\right) \propto \exp \left( \beta E\right) $. This failure comes from the fact that in the $3-D$ case, such entropy maximizer no exists. Nevertheless, interesting features come from such an
analysis when the system is put in an unphysical box $-$ see \cite{P06} for a
review of this topic. Another problem is the fact that the Boltzmann
entropy
\[
S=-\int \;f\;\ln f\;d\mathbf{r}d\mathbf{p}
\]
which is extremalized in this approach, is a conserved Casimir functional in
the Vlasov-Poisson context
\footnote{A Casimir functional is on the form $\mathcal{C}[f]=\int C(f)d\mathbf{r}d\mathbf{p}$ where $C$ is a smooth function. It is well known that such functionals are time conserved quantities if $f$ is a solution of the Vlasov equation:
\begin{eqnarray*}
\frac{d\mathcal{C}[f]}{dt}=\int \frac{d C(f)}{df}\frac{\partial f}{\partial t}d\mathbf{r}d\mathbf{p}&=&
-\int \frac{d C(f)}{df}\left\{ \frac{\mathbf{p}}{m}\frac{\partial f}{\partial \mathbf{x}} - m\frac{\partial \psi}{\partial \mathbf{x}}\frac{\partial f}{\partial \mathbf{p}}\right\}d\mathbf{r}d\mathbf{p} \\
&=& -\int \left\{ \frac{\mathbf{p}}{m}\frac{\partial C}{\partial \mathbf{x}} - m\frac{\partial \psi}{\partial \mathbf{x}}\frac{\partial C}{\partial \mathbf{p}}\right\}d\mathbf{r}d\mathbf{p}=0
\end{eqnarray*}
where the last equality follows by intergrating the first term over $\mathbf{x}$ and the second term over $\mathbf{p}$, since $f\rightarrow0$ as $\left|\mathbf{x}\right|$, $\left|\mathbf{p}\right|\rightarrow\infty$.}
!

Observational approach consists in the integration in the models of observed
properties of globular clusters or galaxies. A non exhaustive list of such a
models is presented in classical text books like \cite{BT87} or more
recently \cite{B00}. Most famous ones are the King model for globular clusters
which is an arbitrarily truncated isothermal sphere, and the
Navarro-Frenk-White\cite{NFW} radial profile for dark matter halos associated to the 
galaxies. From a theoretical point of view, all these models are poorly
justified, except perhaps the Henon isochrone model \cite{HXX}
which is unfortunately generally unrealistic ...

The last way is numerical. Taking into account the recent ability given to
stellar dynamicist by numerical tools to produce realistic formation
experiences, important advances were made at the turning of the last century. These analysis are presented in the next section.

\section{A numerical approach to equilibrium properties}

The idea is to produce an equilibrium state from the gravitational 
collapse of an arbitrary physical set of point masses. This is not a recent idea and it was
pioneered by Spitzer since the late sixties. The effective realization of
such a study in  a general way was the result of a serie of works displayed over more than twenty years (\cite{RP04} and reference therein, \cite{AB06} and reference therein). 
In those works , a general understanding of observed equilibrium properties of globular clusters and dark matter halos of galaxies follows from a detailed analysis of numerical gravitational collapses of a sufficiently large set of $N$ point masses.
Gravitational collapse from physical initial state produces generally a
sphere. Flatness is possible by Radial Orbit Instability but it requires strong and inhomogeneous collapse
\footnote{An important amount of solid rotation could also be invoked but this property is generally not observed in a sufficient way in stellar systems, like elliptical galaxies in particular.}. The radial mass density 
profile of such spheres splits into two distinct classes : Homogeneous
initial conditions collapses into  a core halo-structure. It consists of a
constant density region (the core) which extension can reach the half-mass radius
of the system. This core is surrounded by a radial power law decreasing density  region
(the halo). The other class is composed by sufficiently inhomogeneous initial conditions whose collapse toward a
monotonic radial power law decreasing density without notable central core.
The origin of the collapsed core of such systems could be
explained by the effect of the Antonov Instability discovered in the
thermodynamical context (see \cite{P06} for a review). These two classes could be directly linked to the two classes of Newtonian self-gravitating systems which are globular clusters and galaxies. The classical hierarchical galaxy formation scenario is evidently linked to the inhomogeneous class. It produces collapsed core $-$ which is a favorable process to form the observed and always mysterious super massive black holes $-$ and potential flattening in violent cases. The smaller, quiet, more isolated and then homogeneous case, could be naturally interpreted as the generic globular cluster formation process. This could explain in the same operation, their generic spherical shape, their typical core-halo density profile, and finally their generic lack of intermediate mass black hole in central region\footnote{The small observed amount of solid rotation could explain some observed flatness. Moreover, collapsed core of approximatively 15\% of the galactic globular clusters is due to dissipation effects related to the time problem noted in section \ref{sect2} see \cite{RP04} for details}. The case of spiral galaxies formation and evolution is more specific and the role of dissipative process cannot be neglegted. Perhaps Vlasov equation is no more relevant for the description of such structures.
\section{conclusion}
Taking into account some physical constraints, the globular clusters and the galaxies could be suitably modelized by the Vlasov-Poisson model.
In addition to this modelization, accurate numerical simulations allows to obtain a global understanding of important stages of their evolutions and of their main differences. However, a lot of fundamental works are always to be done before the study of very  particular properties of the self-gravitating systems.

\section*{Acknowledgements}
The author thanks the referees for their very careful reading of the first version of the paper, and M. Kiessling for the communication of the fundamental reference \cite{emden}.

\end{document}